\documentclass[%
 reprint,
superscriptaddress,
amsmath,amssymb,
aps,
prb,
]{revtex4-2}

\usepackage{graphicx}
\usepackage{dcolumn}
\usepackage{bm}%
\usepackage{graphicx,graphics,color,epsfig}
\usepackage{amsmath}
\usepackage{mathrsfs}
\usepackage{amssymb}
\usepackage{subfigure}
\usepackage{amsfonts}
\usepackage{graphicx}
\usepackage{multirow,booktabs}
\usepackage{float}
\begin{document}

\title{Characterizing dynamical behaviors in topological open systems with boundary dissipations}
\author{Zhen-Yu Zheng}
\affiliation{Beijing National Laboratory for Condensed Matter Physics, Institute of Physics, Chinese Academy of Sciences, Beijing 100190, China}
\author{Xueliang Wang}
\affiliation{Beijing National Laboratory for Condensed Matter Physics, Institute of Physics, Chinese Academy of Sciences, Beijing 100190, China}
\affiliation{School of Physical Sciences, University of Chinese Academy of Sciences, Beijing 100049, China}
\author{Shu Chen}
\email{schen@iphy.ac.cn }
\affiliation{Beijing National Laboratory for Condensed Matter Physics, Institute of Physics, Chinese Academy of Sciences, Beijing 100190, China}
\affiliation{School of Physical Sciences, University of Chinese Academy of Sciences, Beijing 100049, China}
\date{\today}

\begin{abstract}
We investigate the dynamics of the Su-Schrieffer-Heeger model with boundary dissipations described by Lindblad master equations  and unravel distinct  dynamical features in the topologically different phases of the underlying Hamiltonian. By examining the long-time damping dynamics, we uncover a dynamical duality phenomenon  between the weak and strong dissipation region,  which exists only in the topologically non-trivial phase, linked to the structure of the Liouvillian spectra, particularly the stripe closest to the steady state. When dissipation is confined to a single boundary, the dynamical duality phenomenon still exists. Under this condition, the Liouvillian gap fulfills an exponential size scaling relation in the topologically non-trivial phase and a power-law size scaling relation in the topologically trivial phase. Within the topologically non-trivial region, we identify the existence of boundary-localized dark states in the thermodynamical limit, which is responsible for the exponential size decay of Liouvillian gap.
\end{abstract}

\maketitle

\section{Introduction}
Advances in quantum engineering of dissipation in laboratory
have attracted a growing interest in the study of the non-equilibrium dynamical features in condensed matter systems with engineered dissipations.
In the realm of open quantum systems, density operators are commonly employed for description of the dynamical evolution of systems, which is typically governed by master equations \cite{breuer2002,Schindler,weimer}. In many studies, a Markovian approximation is employed, allowing the dynamic behavior to be effectively captured by the Lindblad master equation \cite{lindblad1976cmp}. Recently, many theoretical researches have been dedicated to understanding the dynamics arising from boundary dissipation \cite{prosen2009jsm,prosen2008prl,prosen2008njp,prosen2011prl,katsura2019prb,katsura2020ptep,znidaric2010jsm,znidaric2015pre,zhou2020prb,zhoubz2022prb,GuoC2018,Rossini2021prb,
vicari2021prb,landi2022rmp,yamanaka2023scp,Zhai,zheng2023prb,Javed2024arxiv}.

The Su-Schrieffer-Heeger (SSH) model \cite{su1979prl}, despite its simplicity as a quadratic fermionic chain, exhibits a paradigmatic example illustrating the emergence of edge states associated with non-trivial topology \cite{LiLH}. Investigating the dynamic features of topology through such a model is valuable not only for fundamental research \cite{YangC} but also for the potential application in quantum information processing \cite{GongZ,SSH-exp1,LinYC}. 
 Recently, there has been growing interest in  dissipative topological models \cite{song2019prl,kou2023prb,LiuCH,YiXX,zhu2014pra,klett2017pra,dangel2018pra,liu2022prad,song2021prb,klett2018epjd,YangF,Ghosh,Paszko,Diehl,Obuse,Lieu}, revealing intriguing topological structure or dynamical behaviors. Most previous studies on the boundary dissipated SSH model \cite{zhu2014pra,klett2017pra,dangel2018pra,liu2022prad,song2021prb,klett2018epjd} mainly focused on the framework of effective non-Hermitian Hamiltonian by discarding the quantum jump terms. However, neglecting quantum jumps of the dissipative dynamics cannot consistently describe the steady state and suffers the problem of probability conservation of density matrix. Therefore, a complete characterization of dissipative dynamics from the full Liouvillian is necessary for understanding properly the topological effect  of the underlying Hamiltonian on the dynamical behaviors.

In this work, we study the dynamical behaviors in one-dimensional topological lattices with boundary dissipations by considering the SSH model as a concrete example. We shall unravel the different dynamics characteristic of the topological open systems when the underlying Hamiltonian lies on the topologically trivial or nontrivial phases. 
In the scheme of the third quantization method \cite{prosen2008njp,prosen2010jsm,zhang2022JSM,GuoC,wang2025arxiv}, the Liouvillian spectrum can be constructed from the rapidity spectrum. The solution of the Liouvillian spectrum  of the SSH model with boundary dissipation is thus reduced to solving a SSH model with imaginary boundary potentials.
Such a reduction enables us to derive analytical results for the rapidity spectra across the entire parameter space.
The Liouvillian spectrum in the strong and weak dissipation region displays similar structures in the topologically nontrivial phase, and thus the long-time relaxation dynamics exhibits identical behaviors, when the weak and strong dissipation strength fulfills a dual relation. However, this intriguing dynamical dual phenomenon is absent in the topologically trivial phase, which is discernible through the analysis of the rapidity spectra. Furthermore, we unveil that in the presence of dissipation at only one boundary, the Liouvillian gap scales as $e^{- \kappa N}$ in the topologically non-trivial phase and $N^{-3}$ in the topologically trivial phase. The exponential decay of Liouvillian gap can be attributed to the existence of a dark state in the topologically non-trivial phase, which is localized at the other boundary.

The remanning paper is organized as follows: In Section II, we demonstrate how to solve the Liouvillian spectrum of the SSH model with boundary dissipation using the third quantization method. This Liouvillian spectrum can be constructed from the rapidity spectrum, thus simplifying the problem to solving the SSH model with imaginary boundary potentials. In Section III, we analyze the Liouvillian spectral features and damping dynamics behavior in both the underlying topologically non-trivial and trivial phases. We find that regardless of whether the boundary dissipation is loss or gain, in the topologically nontrivial phase, the long-time relaxation dynamics exhibits a duality between the weak and strong dissipation regimes, whereas no such a duality exists in the topologically trivial phase. This dynamical duality phenomenon can be analyzed from either the Liouvillian spectrum or the rapidity spectrum. In Section IV, we focus on the case where only one boundary with dissipation. We explore the Liouvillian gap and its scaling, pointing out the presence of dark states in the topologically nontrivial phase. We find that the Liouvillian gap fulfills an exponential size scaling  $e^{- \kappa N}$ in the topologically non-trivial phase and a power-law size scaling $N^{-3}$ in the topologically trivial phase, and investigate the effect of these dark states on the long-time dynamics. A summary is given in the final section.

\section{Model and formalism}


We start with the SSH model with open boundary condition by $2N$ sites arranged in $N$ unit cells and the Hamiltonian of model is given by
\begin{align}\label{SSH}
 H=\sum_{j=1}^{N}\left(t_{1} c_{2 j-1}^{\dagger} c_{2 j}+t_{2} c_{2 j}^{\dagger} c_{2 j+1}+\text { H.c. }\right),
\end{align}
where $c_{j}^{\dagger}, c_{j}$ are fermion creation and annihilation operators at site $j$, $t_{1}$ and $t_{2}$ are hopping amplitudes between intra-cell and inter-cell sites, and the summation of $j$ is over $N$ cells.   By tuning $t_{1}$, the model carries out a phase transition from the topologically nontrivial phase ($t_1<t_2$) to topologically trivial phase ($t_1>t_2$) with the phase transition point occurring at $t_{1}/t_{2}=1$. For convenience, we shall set $t_{2}=1$ as the energy unit in the following calculation.

Within the Markovian approximation, the dynamic evolution process of a boundary-driven quantum system is governed by the Lindblad master equation:
\begin{equation}\label{lindblad}
  \frac{d\rho}{dt}=\mathcal{L}[\rho]:=-\mathrm{i}[H,\rho]+\sum_{i}(2L_{i}\rho L_{i}^{\dagger}-\{L_{i}^{\dagger}L_{i},\rho\}),
\end{equation}
where $\rho$ is the density matrix and we have set $\hbar=1$.
Here, we consider the quantum jumps due to the environment appearing at left and right boundaries which are described by the Lindblad loss or gain dissipators $L_{l/r}^{l/g}$, i.e.
\begin{equation}\label{LLO}
  L_{l/r}^{l}=\sqrt{\gamma_{l/r}}c_{l/r}, ~~ L_{l/r}^{g}=\sqrt{\gamma_{l/r}}c_{l/r}^{\dagger},
\end{equation}
where $\gamma_{l},\gamma_{r}$ denote the dissipation strengths on left and right boundary, respectively.

The SSH model with boundary dissipations can be solved by the third quantization method \cite{prosen2008njp,prosen2010jsm,zhang2022JSM,GuoC,wang2025arxiv}. According to this method, we firstly find a set of Majorana fermion operators as follow,
\begin{align}
\left(\begin{array}{c}w_{2j-1} \\ w_{2j}\end{array}\right):=\left(\begin{array}{cc}1 & 1 \\ i & -i\end{array}\right)\left(\begin{array}{c}c_j \\ c_j^{\dagger}\end{array}\right),
\end{align}
and they obey the anti-commutation relations $\{w_{i},w_{j}\}=2\delta_{i,j}$.

Then, the SSH Hamiltonian and Lindblad operators can be expressed in
terms of a quadratic form and linear forms in Majorana fermion form, respectively,
\begin{equation}\label{qflf}
  \begin{aligned}
H & =\sum_{j, k=1}^{4N} w_j \mathbf{H}_{j, k} w_k=\underline{w} \cdot \mathbf{H} \underline{w}, \\
L_\mu & =\sum_{j=1}^{N} l_{\mu, j} w_j=\underline{l_\mu} \cdot \underline{w},
\end{aligned}
\end{equation}
where $\underline{w}=\left(w_1, w_2, \cdots\right)^{\mathrm{T}}$ will designate a vector (column) of appropriate scalar valued or operator valued symbols $w_k$. Here, we should notice that the $4 N \times 4 N$ matrix $\mathbf{H}$ can be chosen to be an anti-symmetric matrix which satisfies $\mathbf{H}^{\mathrm{T}}=-\mathbf{H}$. Next, we discuss the matrix of the Lindblad dissipative operators in Majorana fermion form defined as
\begin{equation}\label{MM}
  \mathbf{M}_{j, k}=\sum_\mu l_{\mu, j} l_{\mu, k}^*.
\end{equation}
The matrix $\mathbf{M}$ is Hermitian matrix. $\mathbf{M}_\mathbf{R}$ and $\mathbf{M}_\mathbf{I}$ are real part and imaginary part of the matrix $\mathbf{M}$, respectively.

By using a Liouville-Fock space of operators\cite{prosen2010jsm}, the Liouvillian can be expressed in the Liouville-Fock picture as
\begin{equation}
\hat{\mathcal{L}}=\hat{\underline{w}} \mathbf{A} \hat{\underline{w}}-A_{0}\hat{\mathbf{1}},
\end{equation}
where $A_{0}=2\text{Tr}(\mathbf{M}_\mathbf{R})$ and $\hat{\underline{w}}$ is the superoperators with $\{\hat{w}_{i},\hat{w}_{j}\}=\delta_{i,j}$. The Liouville structure matrix $\mathbf{A}$ is a $8N\times8N$ antisymmetric matrix with the following form:
\begin{equation}\label{Am}
\mathbf{A}=\left(\begin{array}{cc}
-2 \mathrm{i} \mathbf{H}+2\mathrm{i} \mathbf{M}_\mathbf{I} & 2 \mathrm{i} \mathbf{M} \\
-2 \mathrm{i} \mathbf{M}^{\mathrm{T}} & -2 \mathrm{i} \mathbf{H}-2\mathrm{i} \mathbf{M}_\mathbf{I}
\end{array}\right).
\end{equation}
In order to exactly solve the spectra of the Liouvillean $\hat{\mathcal{L}}$, we can rewrite the Liouville structure matrix $\mathbf{A}$ as a block-triangular matrix $\tilde{\mathbf{A}}$:
\begin{equation}\label{tildeA}
\tilde{\mathbf{A}}=\mathbf{U}\mathbf{A}\mathbf{U}^{\dagger}=\left(\begin{array}{cc}
-\mathbf{X}^{T} & 4 \mathrm{i} \mathbf{M}_\mathbf{I} \\
0 & \mathbf{X}
\end{array}\right),
\end{equation}
where $\mathbf{U}=\frac{1}{\sqrt{2}} \left(\begin{array}{cc}
1 & -\mathrm{i} \\
1 & \mathrm{i}
\end{array}\right) \otimes \mathbf{1}_{4N}$ and $\mathbf{X}=-2 \mathrm{i}\mathbf{H}+2 \mathbf{M}_\mathbf{R}$.
It is well-established that the characteristic polynomial of the structure matrix $\mathbf{A}$ factorizes in terms of the characteristic polynomial of $\mathbf{X}$\cite{prosen2010jsm}. Additionally, the Liouvillian $\hat{\mathcal{L}}$ can be constructed based on the eigenvalues of the $4N\times4N$ matrix $\mathbf{X}$.

Let us delve into the properties of the matrix $\mathbf{X}$. We consider a basis transformation to change the matrix $\tilde{\mathbf{A}}$ into a block-diagonal form:
\begin{equation}
\tilde{\mathbf{A}}=\mathbf{W}^{-1}\left(\begin{array}{cc}
-\mathbf{X}^{T} & 0 \\
0 & \mathbf{X}
\end{array}\right)\mathbf{W},
\end{equation}
where $\mathbf{W}=\left(\begin{array}{cc}
\mathbf{I} & -4 \mathrm{i} \mathbf{Z} \\
0 & \mathbf{I}
\end{array}\right)$. The $\mathbf{Z}$ is a solution of the continuous Lyapunov equation:
\begin{equation}
\mathbf{X}^{T}\mathbf{Z}+\mathbf{Z}\mathbf{X}=\mathbf{M}_\mathbf{I}.
\end{equation}


Assuming the matrix $\mathbf{X}$ can be diagonalized as follows:
\begin{equation}
\mathbf{X}=\mathbf{V}^{-1} \mathbf{R} \mathbf{V},
\end{equation}
where the matrix $\mathbf{R}$ is a diagonal matrix. And we define:
\begin{equation}
\left(\begin{array}{cc}
\hat{g}_j  \\
\hat{g}_{j+4N}^{\prime}
\end{array}\right)= \left(\begin{array}{cc}
(\mathbf{V}^{T})^{-1} & 0 \\
0 & \mathbf{V}
\end{array}\right)\mathbf{W} \mathbf{U} \left(\begin{array}{cc}
\hat{w}_{2j-1}  \\
\hat{w}_{2j}
\end{array}\right),
\end{equation}
where $\hat{g}_j$ and $\hat{g}'_j$ satisfy almost-canonical commutation relation as
$
\left\{\hat{g}_j, \hat{g}_k\right\}=\left\{\hat{g}_j^{\prime}, \hat{g}_k^{\prime}\right\}=0,\left\{\hat{g}_j, \hat{g}_k^{\prime}\right\}=\delta_{j k}$.

Thus, the Liouvillean $\hat{\mathcal{L}}$ can be rewrite as:
\begin{equation}
\hat{\mathcal{L}}=-2 \sum_{j=1}^{4 N} \alpha_{j} \hat{g}_j^{\prime} \hat{g}_j.
\end{equation}
Here $\alpha_{j}$ are the spectra of the matrix $\mathbf{X}$ and the form of matrix $\mathbf{X}$ can be represented as:
\begin{equation}\label{x}
\mathbf{X}=-\frac{1}{2}\mathbf{S}^{-1}\left(\begin{array}{cc}
\mathrm{i} \mathbf{P}(t_{1}, t_{2}, \gamma_{l}, \gamma_{r}) & 0 \\
0 & \mathrm{i} \mathbf{P}(-t_{1}, -t_{2}, \gamma_{l}, \gamma_{r})
\end{array}\right)\mathbf{S},
\end{equation}
where the matrix $\mathbf{S}=\frac{1}{\sqrt{2}}[\mathbf{1}_{2N}\otimes \left(\begin{array}{cc}
1  \\
-\mathrm{i}
\end{array}\right), \mathbf{1}_{2N}\otimes \left(\begin{array}{cc}
1  \\
\mathrm{i}
\end{array}\right)]$. The spectra $\alpha_{j}$ is determined by the $2N\times2N$ matrix $\mathbf{P}$ given by
\begin{align}
\mathbf{P}(t_{1}, t_{2}, \gamma_{l}, \gamma_{r})=\left(\begin{array}{cccccc}
\mathrm{i} \gamma_{l} & t_{1} &  &  &  & \\
t_{1} & 0 & t_{2}\\
 & t_{2} & 0 & \ddots\\
 &  & \ddots & \ddots & \ddots\\
 &  &  & \ddots & 0 & t_{1}\\
 &  &  &  & t_{1} & \mathrm{i} \gamma_{r}
\end{array}\right).\label{P}
\end{align}
The eigenvalues $E_p$ of the matrix $\mathbf{P}(t_{1}, t_{2}, \gamma_{l}, \gamma_{r})$ and the eigenvalues $E_p'$ of the matrix $\mathbf{P}(-t_{1}, -t_{2}, \gamma_{l}, \gamma_{r})$ collectively referred to as the rapidity spectra, and we have $i E_{j} = - 2\alpha_{j} , (E_{j} \in {E_p \cup E_p'})$.
The Liouvillian spectra $\lambda$ can be constructed by using the rapidity spectra:
\begin{equation}\label{lambda}
\lambda= \mathrm{i} \sum_{j=1}^{4N} v_{j} E_{j}~~(v_{j}=0,1; E_{j} \in {E_{p} \cup E_{p}'}).
\end{equation}

We note that the matrix $\mathbf{P}(t_{1}, t_{2}, \gamma_{l}, \gamma_{r})$ remains unchanged, regardless of whether we select dissipators $L_{l/r}^{l/g}$ with boundary loss or gain. This is because the matrix $\mathbf{X}$, which determines the rapidity spectra, is only relevant to the real part of $M$ matrix and keeps unchanged for either loss or gain dissipators (see Appendix A for details). We also show the details of exactly solving the eigenvalues of matrix $\mathbf{P}(t_{1}, t_{2}, \gamma_{l}, \gamma_{r})$ and prove that the spectra $E_{p}$ are the same as the spectra $E_{p}'$ in Appendix B.  According to Eqs. (\ref{x}) and (\ref{P}), the rapidity spectra do not exhibit $\mathcal{P}\mathcal{T}$ symmetry but instead possess $\mathcal{K}$ symmetry \cite{zhu2014pra,zheng2023prb}, even we choose a balanced boundary loss and gain dissipator.

\section{System with both left and right boundary dissipations}

\subsection{Characters of Liouviallian spectra}
\begin{figure}[h]
\centering \includegraphics[width=8.5cm]{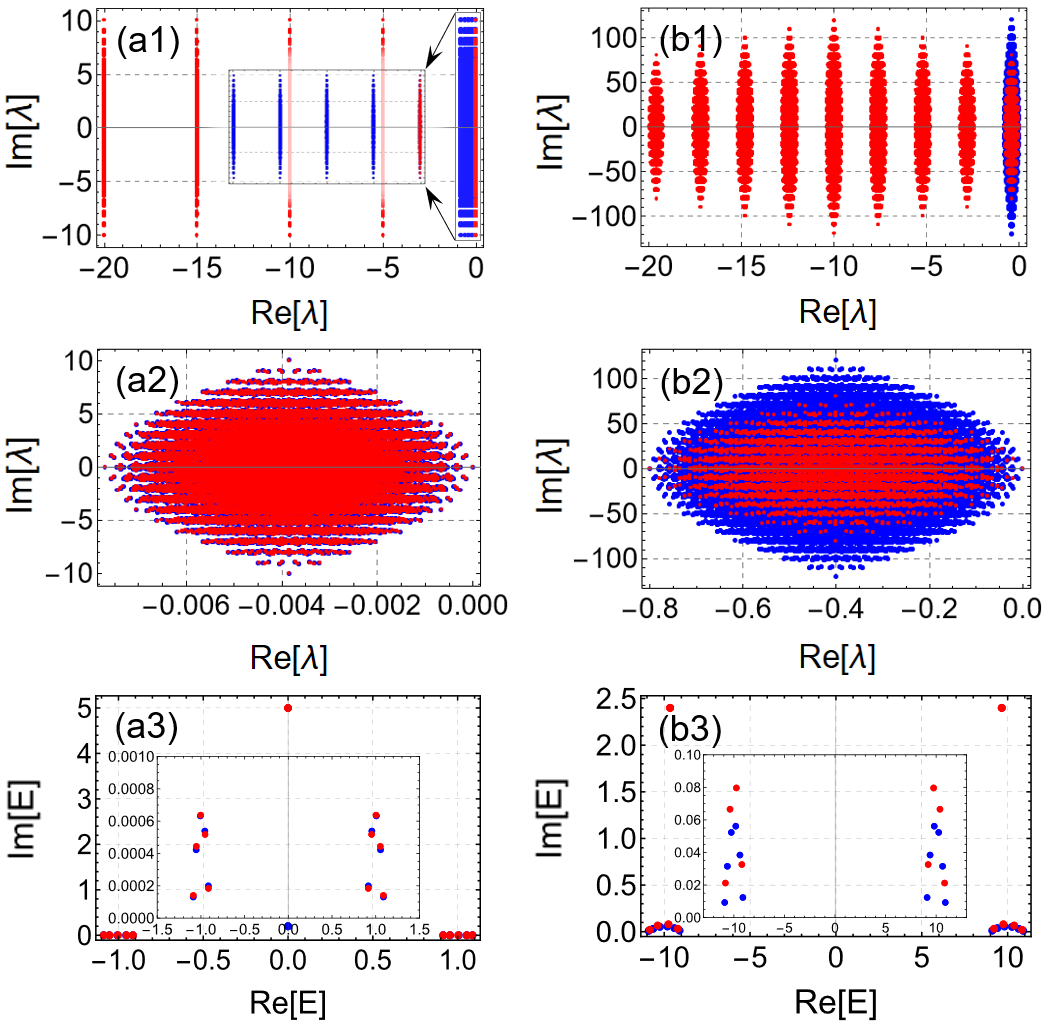} \caption{The Liouvillian spectra, the stripe of Liouvillian spectra closest to the steady state and rapidity spectra with $N=6$, and (a1, a2, a3) $t_{1} =0.1, t_{2}=1$ or (b1, b2, b3) $t_{1} =10, t_{2}=1$. The blue points represent the Liouvillian spectra of the parameters with $\gamma_{l}=\gamma_{r}=0.2$ while the red ones represent that of the parameters with $\gamma_{l}=\gamma_{r}=5$. The eigenvalues of Liouvillian spectra in all panels satisfy $\Re[\lambda]\leq 0$ and the data of Liouvillian spectra are consistent with the ones by exact diagonalization.}
\label{fig1}
\end{figure}

For convenience, we consider a system with boundary dissipators characterized by pure loss ($L_{l/r}^{l}=\sqrt{\gamma_{l/r}}c_{l/r}$).
In Fig.\ref{fig1}(a1) and (b1), we display the Liouvillian spectrum for the topologically nontrivial  $(t_{1}=0.1,t_{2}=1)$ and trivial system $(t_{1}=10,t_{2}=1)$, respectively. While the spectrum for the case with a weak dissipation $\gamma_{l}=\gamma_{r}=0.2$ is marked by blue dots, the spectrum with a strong dissipation $\gamma_{l}=\gamma_{r}=5$ is denoted by red dots.
Here the Liouvillian spectra are constructed by using Eq.(\ref{lambda}), and we have verified their accuracy by comparing them with numerical results obtained through exact diagonalization of the Liouvillian, confirming their exact equivalence.
From Fig.\ref{fig1} (a1) and (b1),  we observe that the Liouvillian spectra display structures of stripes.
For the topologically nontrivial system, the number of stripes for the case with weak dissipation is the same as the case with strong dissipation. In contrast, for the topologically trivial system, the number of stripes varies for the case with weak and strong dissipations. While the spectrum is composed of nine stripes in the strong dissipation region, it only includes one stripe in the weak dissipation region.

The long time dynamics is governed by the Liouvillian modes with Re$(\lambda)$ closest to zero. To see clearly the structure of Liouvillian spectrum  closest to steady state,  we illustrate the rightmost stripe of the Liouvillian spectrum for the topologically nontrivial and trivial systems in  Fig. \ref{fig1}(a2) and Fig. \ref{fig1}(b2), respectively.  For the topologically nontrivial system, it is shown that the spectrum of the rightmost stripe is similar for the weak and strong dissipation case when $\gamma_{week}=1/\gamma_{strong}$ is fulfilled. For the topologically trivial system, the spectrum of the rightmost stripe in the strong dissipation region only overlaps with part of the spectrum in the weak dissipation region even we take  $\gamma_{week}=1/\gamma_{strong}$.

Since the Liouvillian spectrum is constructed from the rapidity spectra by using Eq.(\ref{lambda}), we can understand the structure of Liouvillian spectrum from the structure of rapidity spectra. Similar to quantum Ising model with boundary dissipations \cite{zheng2023prb}, here the different stripe structure of Liouvillian spectrum is determined by the number of boundary bound states in the rapidity spectrum.
As illustrated in Fig. \ref{fig1}(a3) for rapidity spectra of the topologically nontrivial system, there are two degenerate boundary bound states in both weak and strong dissipation regions. Excluding the boundary bound states, the rest of rapidity spectra are similar. This phenomenon is absent for the topologically trivial system. As shown in Fig. \ref{fig1} (b3),  while there are four boundary bound states (two group of degenerate bound states) in the strong dissipation region, no boundary bound state exists in the weak dissipation region.
The distinct number of boundary bound states results in distinct stripe structure in the Liouvillian spectra in the weak and strong dissipation regions. Moreover, the rapidity spectra excluded the boundary-bound states differ in the weak and strong dissipation regions, leading to discrepancies of Liouvillian spectrum in the stripe closest to zero.

\subsection{Dynamical duality in the topologically nontrivial region}
To demonstrate the dynamical evolution, we focus on the evolution of the particle number density  given by \cite{zhang2022JSM}:
\begin{equation}\label{eq:spc}
 n(t)=\frac{1}{2N}\sum_{i=1}^{2N}\text{Tr}[c_{i}^{\dagger}c_{i}\rho(t)].
\end{equation}
For convenience, we choose the fully occupied state as the initial state, i.e., $n(0)=1$. Details of calculation of $n(t)$ can be found in Appendix C.

\begin{figure}
\centering \includegraphics[width=8.5cm]{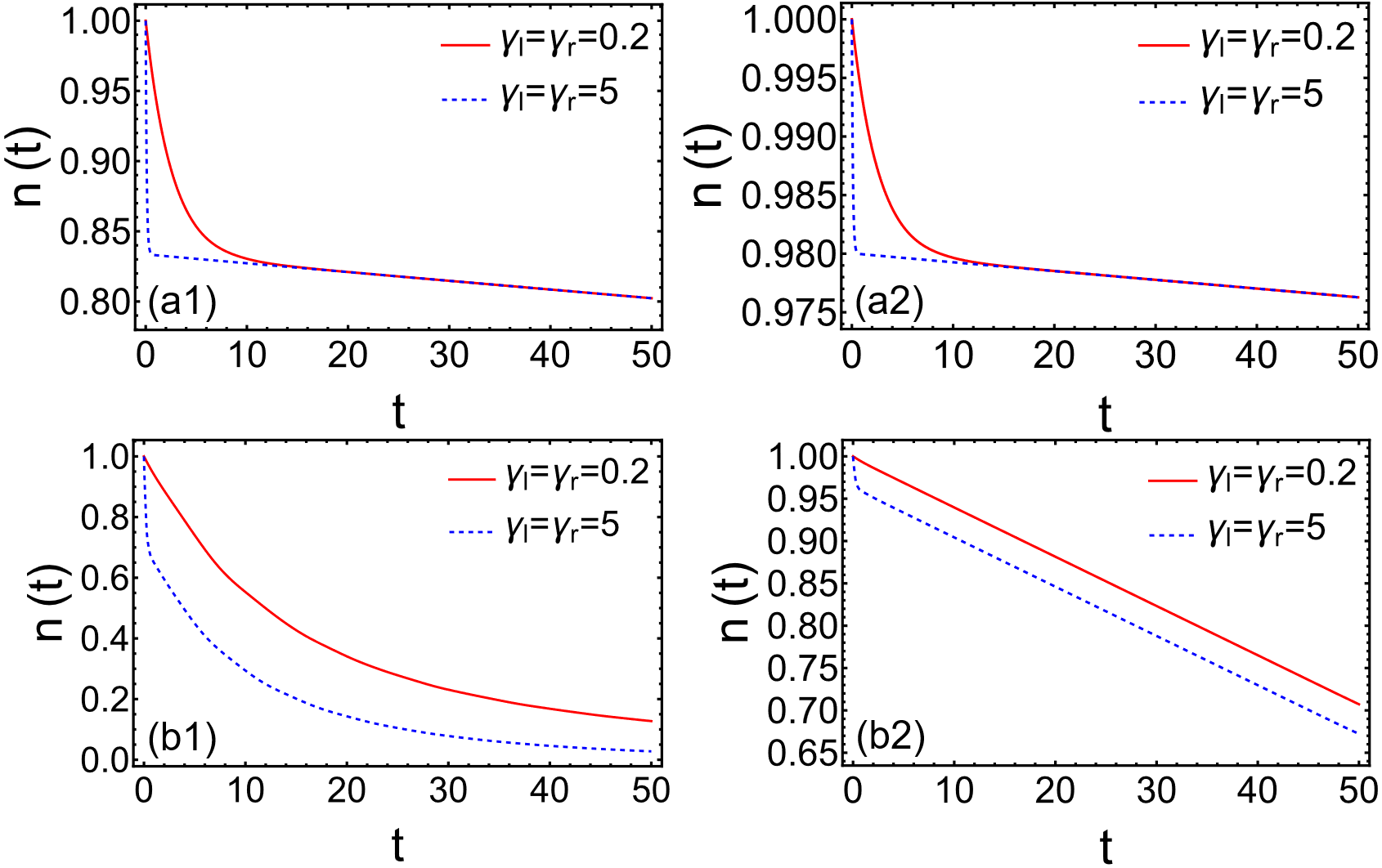} \caption{The dynamical evolution of  the particle number density  with (a1) $N=6$, $t_{1}=0.1$ and $t_{2}=1$, and (a2) $N=50$, $t_{1}=0.1$ and $t_{2}=1$, and (b1) $N=6$, $t_{1}=10$ and $t_{2}=1$, and (b2) $N=50$, $t_{1}=10$ and $t_{2}=1$. Both left and right boundaries are loss.}
\label{fig2}
\end{figure}

In Fig.\ref{fig2}, we present the time evolution of the particle number density for various lattice sizes with the same parameters as in Fig.\ref{fig1}. It is shown that the topologically nontrivial systems with $\gamma=0.2$ and $5$ exhibit almost identical dynamical behaviors in the long time scale, whereas difference only appears in the short time scale, as displayed in Fig.\ref{fig2}(a1) and (a2).
However, for the topologically trivial system with the same dissipation parameters, the dynamical behavior is obviously different in either long or short time scales, as depicted in Fig.\ref{fig2}(b1) and (b2).
The occurrence of identical dynamical behaviors in the long time scale for systems with $\gamma$ and $\gamma'=1/\gamma$ was termed as dynamical duality \cite{zheng2023prb}, which can be observed only in  the topologically nontrivial region of the SSH model with boundary dissipations.

\begin{figure}
\centering \includegraphics[width=8.5cm]{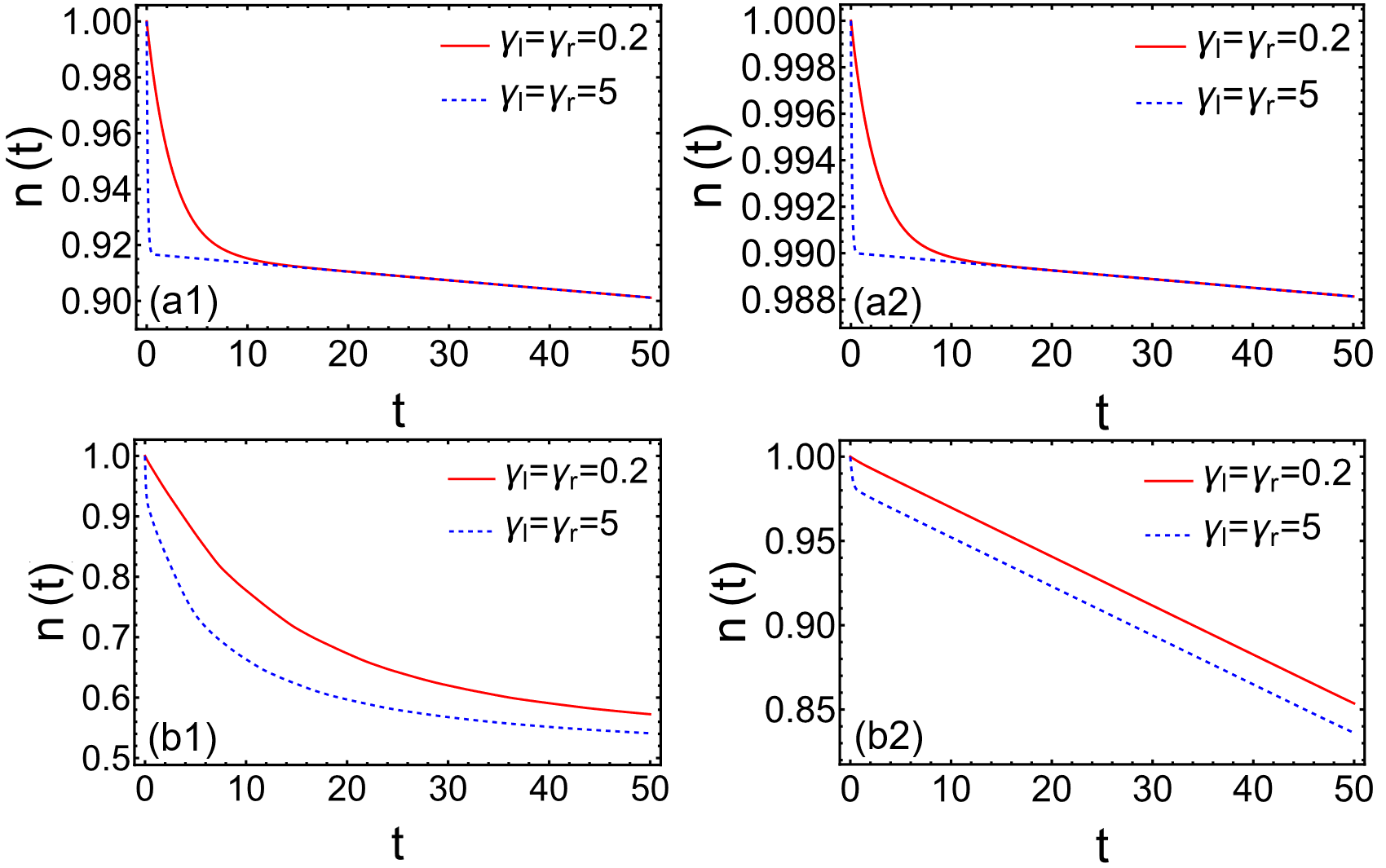} \caption{The dynamical evolution of the particle number density  with (a1) $N=6$, $t_{1}=0.1$ and $t_{2}=1$, and (a2) $N=50$, $t_{1}=0.1$ and $t_{2}=1$, and (b1) $N=6$, $t_{1}=10$ and $t_{2}=1$, and (b2) $N=50$, $t_{1}=10$ and $t_{2}=1$. The left boundary is loss and the right boundary is gain.}
\label{fig3}
\end{figure}
The phenomena of dynamical duality can be understood from the properties of Liouviallian spectrum. While the short-time dynamics is primarily influenced by the leftmost stripes of the Liouvillian spectrum, the long-time dynamics is governed by the rightmost stripe of the Liouvillian spectrum. As the real part of the leftmost stripe of the Liouvillian spectrum with strong dissipation is much larger than that with weak dissipation, $n(t)$ decays more quickly in the short time scale for system with strong dissipation. Since the rightmost stripe of the Liouvillian spectra with $\gamma_{l}=\gamma_{r}=0.2$ and $\gamma_{l}=\gamma_{r}=5$ is almost identical as shown in Fig.\ref{fig1}(a2), we can infer that the long-time relaxation dynamics exhibits dynamical duality behaviors for the topologically nontrivial system.   The dynamical duality is not observed in the topologically trivial region, due to the mismatch of the Liouvillian spectra in the rightmost stripe, as displayed in Fig.\ref{fig1}(b2).  When the lattice size increases, the rapidity spectra with boundary bound states excluded agree better for topologically nontrivial system with weak ($\gamma$) and strong dissipation ($1/\gamma$), which suggests the dynamical duality manifests in the large size limit. In Fig.\ref{fig2}(a2), we show an example of a larger lattice with $N=50$ and can observe consistent dynamical results with those obtained in system with $N=6$. An analytical analysis of the duality relation of rapidity spectrum for the system with dissipation $\gamma$ and $\gamma'=1/\gamma$ is given in Appendix D.

If we change the boundary dissipation operator to gain operators, we find that the Liouvillian spectrum and the rapidity spectrum remain unchanged, as shown in Fig.\ref{fig1}, although the steady state changes (see Appendix C). Since the Liouvillian spectrum does not change, we can conclude that the long-time dynamical behavior still exhibits dynamical duality in the topologically nontrivial region, while no such a duality occurs in the topologically trivial region. This is confirmed by the numerical results shown in Fig.\ref{fig3}, displaying the dynamical behaviors when one boundary dissipation is changed to a gain.

\begin{figure}
\centering \includegraphics[width=8.5cm]{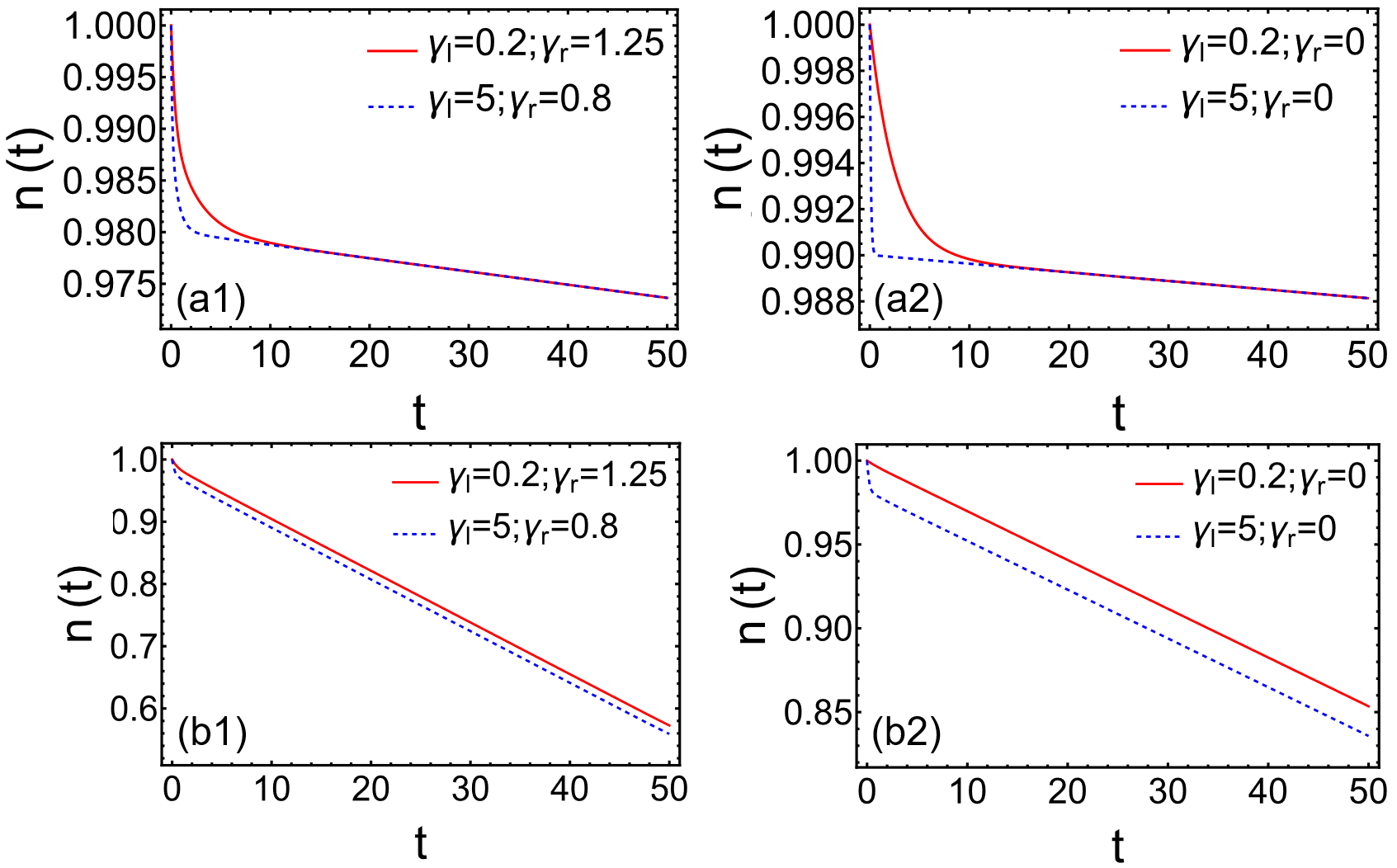} \caption{The dynamical evolution of  the particle number density  with (a1, a2) $N=50$, $t_{1}=0.1$, $t_{2}=1$, and (b1, b2) $N=50$, $t_{1}=10$, $t_{2}=1$.}
\label{fig4}
\end{figure}

The dynamical duality also emerges in systems with different dissipation strengths at the left and right boundaries, as long as  $\gamma'_l=1/\gamma_l$ and $\gamma'_r=1/\gamma_r$ are fulfilled separately. In Fig.\ref{fig4}(a1) and (b1), we adjust these dissipation strengths while maintaining the relation between weak and strong dissipation. We observe that the long time dynamics exhibits dynamical duality in the topologically nontrivial region, whereas this behavior is absent in the topologically trivial region. In Fig.\ref{fig4}(a2) and (b2), we consider cases with dissipation applied only at the left boundary. It is shown that the long time dynamics still demonstrates dynamical duality in the topologically nontrivial region but not in the trivial region.

\section{System with single boundary dissipator}
In this section, we consider the case with dissipation applied only at one of the boundary. Here, we mainly investigate the case of $\gamma_{l}=\gamma, \gamma_{r}=0$ and the case of $\gamma_{l}=0, \gamma_{r}=\gamma$ can be obtained with mirror symmetry.
We shall calculate the Liouvillian gap $\Delta_{g}$, which is defined as the spectral gap between the first decay modes and the steady state: \cite{znidaric2015pre,CaiZ}
\begin{equation}\label{gap}
  \Delta_{g}:=-  \max ~ \Re(\lambda)|_{\Re(\lambda)\neq0}.
\end{equation}
The Liouvillian gap provides important information for the relaxation time of the quantum dissipative system.
\begin{figure}[h]
\centering \includegraphics[width=8.5cm]{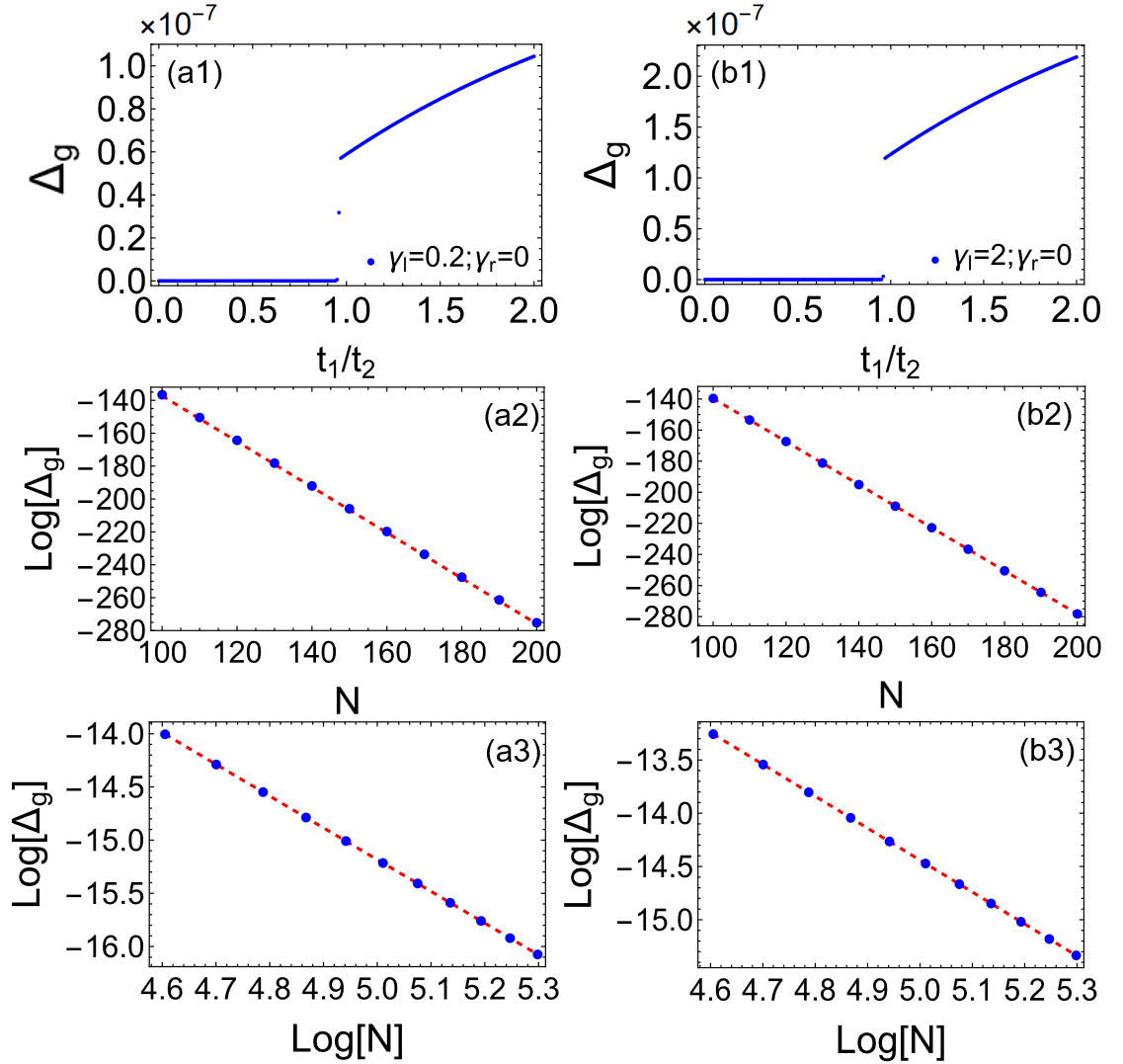} \caption{The Liouvillian gap with $N=200$ in (a1, a2). Finite size scaling of the Liouvillian gap with (a2) $t_{1}=0.5, t_{2}=1, \gamma=0.2$ and (b2) $t_{1}=0.5, t_{2}=1, \gamma=2$ and (a3) $t_{1}=2, t_{2}=1, \gamma=0.2$ and (b3) $t_{1}=2, t_{2}=1, \gamma=2$. The blue points are the data by strict diagonalization and the red dashed lines are the fit data.}
\label{fig5}
\end{figure}

In Fig. \ref{fig5} (a1) and (b1), we illustrate the variation of the Liouvillian gap with  $t_{1}$ under weak and strong dissipation, respectively. The Liouvillian gap undergoes a sharp change at the point of $t_{1}/t_{2}=1$, irrespective of whether it is under strong or weak dissipation. This point is the same as the topologically phase transition point in the SSH model.
By analyzing the finite size scaling of the Liouvillian gap, we find that the Liouvillian gap can be well fitted as: $\Delta_{g}=3.749\times\mathrm{e}^{-1.3863N}$ for Fig.\ref{fig5} (a2), $\Delta_{g}=0.7772\times N^{-2.9863}$ for Fig.\ref{fig5} (a3), $\Delta_{g}=0.3749\times\mathrm{e}^{-1.3863N}$ for Fig.\ref{fig5} (b2), and $\Delta_{g}=1.7247\times N^{-2.9971}$ for Fig.\ref{fig5} (b3).
These results reveal that the Liouvillian gap in the topologically nontrivial region fulfills an exponential scaling relation $\Delta_{g}\thicksim\mathrm{e}^{- \kappa N}$ where $\kappa$ is a non-universal constant, while in the topologically trivial region, it displays a power-law size scaling  $\Delta_{g}\thicksim  N^{-3}$.


\begin{figure}[h]
\centering \includegraphics[width=8.5cm]{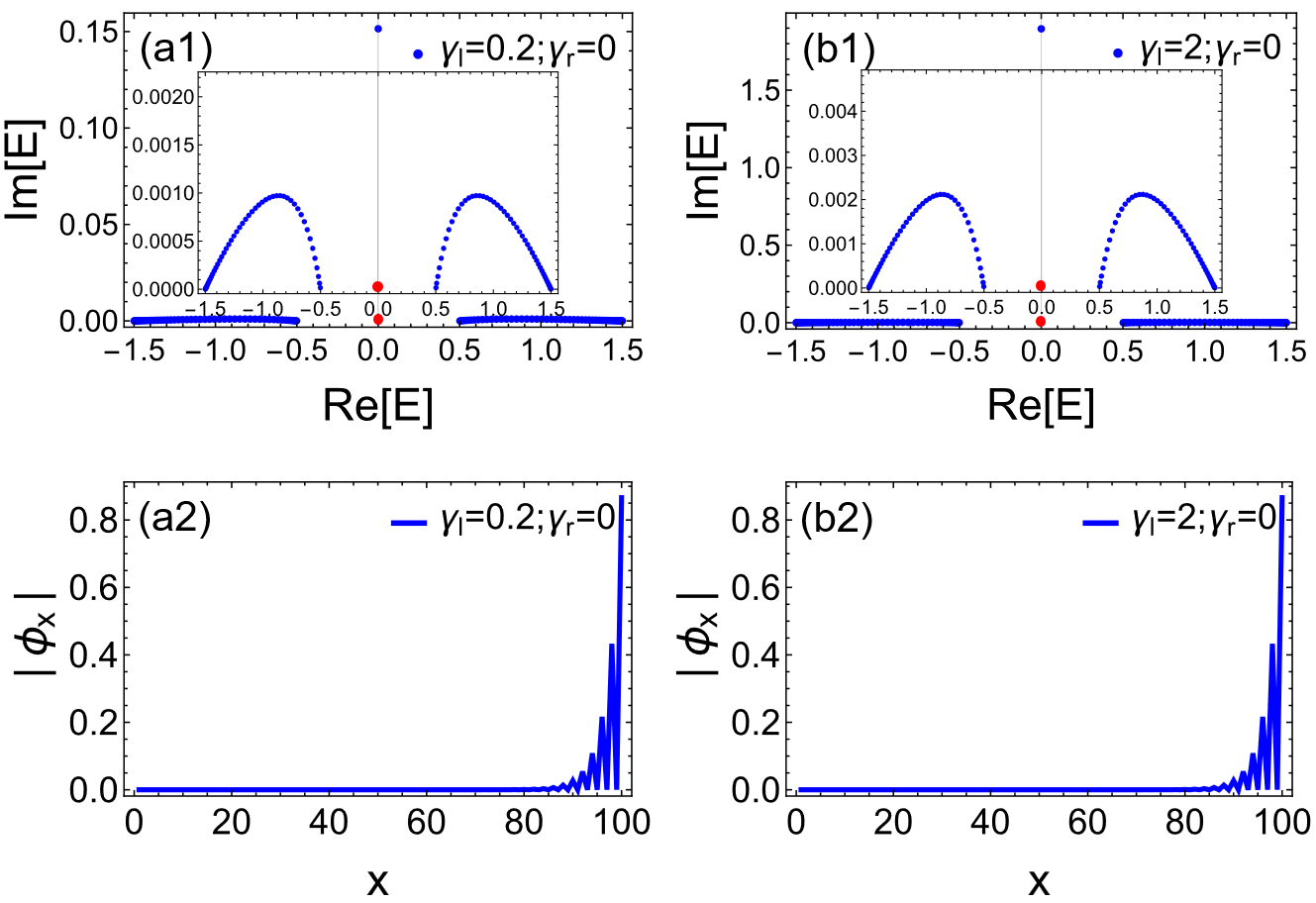} \caption{The rapidity spectra with $N=50, t_{1} =0.5, t_{2}=1$, (a1) $\gamma_{l}=0.2, \gamma_{r}=0$ and (b1) $\gamma_{l}=2, \gamma_{r}=0$. (a2) and (b2) show the wave function of the dark state, indicated by a red point in (a1) and (b1), respectively. The $\mathbf{x}$ is the site index of the unit cell.}
\label{fig6}
\end{figure}

The occurrence of an exponentially small Liouvillian gap can be attributed to the emergence of topological edge states in the topologically nontrivial region. When a boundary dissipation is applied at one of the edge, the edge state localized on the other boundary contributes a near-zero mode to the rapidity spectrum. Such a mode is exactly a zero mode only in the thermodynamical limit, corresponding to a dark sate of the dissipated SSH model \cite{kou2023prb}.

In Fig.\ref{fig6}, we illustrate the dark states within the rapidity spectra in the topologically nontrivial phase, marked by red points, along with the corresponding wavefunctions. The dark state in this spectrum is characterized by both its real and imaginary parts being approximately zero. In the thermodynamic limit, it can be rigorously shown that the spectrum of the dark state exactly reaches zero, as detailed in Appendix E. Regardless of the dissipation strength, whether strong or weak, the wavefunction distribution of the dark state remains localized at the right boundary. Furthermore, the distribution of the dark state wavefunction is independent of the dissipation strength but depends on the hopping amplitudes between intra-cell and inter-cell sites, as discussed in Appendix E. This boundary localization significantly influences the dynamic behavior and leads to an exponential-size decay of Liouvillian gap \cite{zhoubz2022prb}.

\begin{figure}
\centering \includegraphics[width=8.5cm]{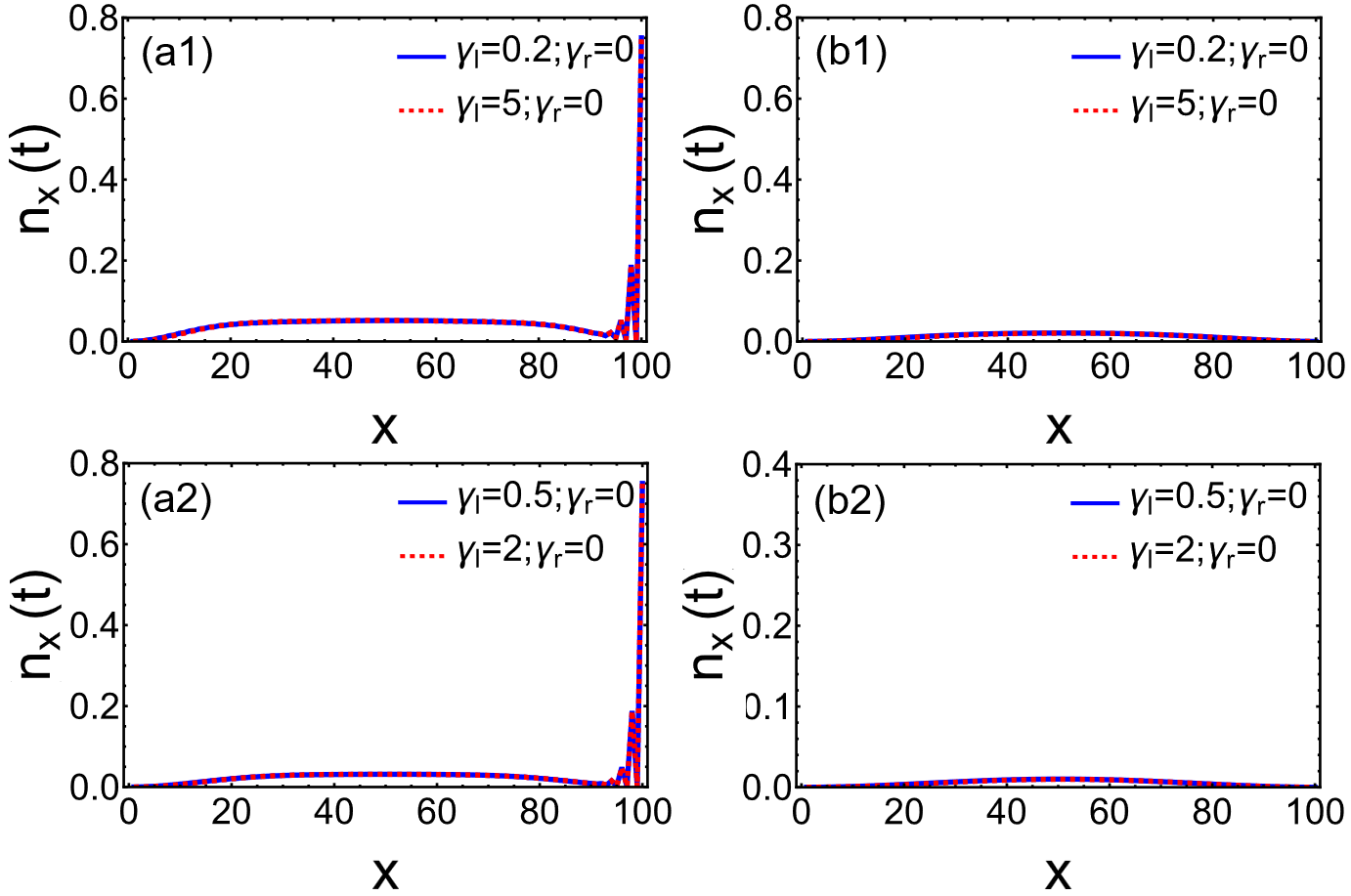} \caption{Particle number distribution in long time. The parameters of the calculation is $N = 50$, (a1), (a2) $t_1 = 0.5$, $t_2=1$, and (b1), (b2) $t_1 = 2$, $t_2=1$, and the time for the evolution is $t = 50000$.}
\label{fig7}
\end{figure}

To better observe the influence of the dark state on the dynamical behavior, we present the particle number distribution $n_x(t)=\text{Tr}[c_{x}^{\dagger}c_{x}\rho(t)]$ after a long evolution time, illustrating the stark contrast between the topologically nontrivial and trivial phases in Fig.\ref{fig7}. In the topologically nontrivial phase as shown in Fig.\ref{fig7}(a1) and (a2), the presence of dark states leads to a localized particle number distribution at the right boundary. The imaginary part of the eigenvalue of the dark state approaches zero in the  thermodynamic limit, ensuring exponentially slow decay of this boundary state and thus sustaining the boundary localization over time. In contrast, within the topologically trivial phase in Fig.\ref{fig7}(b1) and (b2), no dark state exists and the particle number distribution spreads uniformly across the entire lattice. 

\section{Conclusion and discussion}
In summary, we studied the dynamics of a SSH model with boundary dissipations in the framework of Lindblad master equations. Our focus lies on elucidating distinct  dynamical features of the topological open system in the topologically non-trivial and trivial phases of the underlying Hamiltonian. We scrutinized the properties of the Liouvillian spectra and rapidity spectra and unveiled that the Liouvillian spectra in the strong and weak dissipation region displays similar structures in the topologically nontrivial phase, which leads to
the existence of a duality relation in long-time relaxation dynamics between weak and strong dissipation regimes. However, the dynamical duality is absent if the underlying Hamiltonian is in the topologically trivial phase.
Furthermore, we studied the scaling behavior of the Liouvillian gap under dissipation at a single boundary, showcasing exponential size scaling in the topologically non-trivial region and power-law size scaling in the trivial region. The exponential decay behavior of Liouvillian gap is closely related to the existence of a dark state localized at the other boundary in the non-trivial phase. Our study deepens our comprehension of dynamics in topological open systems, thereby furnishing insights for experimental inquiries.

In this work, we focused on the topological systems with boundary dissipations. A typical feature of the underlying topological system is the existence of localized boundary modes in the topological phase, in which we observe the phenomenon of dynamical duality. It is interesting to ask whether dynamical duality behavior can also emerge in non-topological systems when localized boundary states are induced by adding boundary defects? To illustrate this, we consider
a simple one-dimensional chain with boundary defects (see Appendix F for details). Our calculation demonstrates the existence of the dynamical duality when the strength of boundary defects exceeds a threshold and induces boundary states. In contrast, when the strength of boundary defects is small than the threshold, no boundary states are generated. In this case, we do not observe the  phenomenon of dynamical duality. Similarly, the difference of long time dynamical behaviors can also be understood from the  similarity of the structure of Liouvillian spectra in the strong and weak dissipation region. An interesting issue is the universality of the phenomenon of dynamical duality, which is worthy of further study in the future work.

\begin{acknowledgments}
We thank X. D. Liu and C. G. Liang for useful discussions. The work is supported by National Key Research and Development Program
of China (Grant No. 2021YFA1402104), the NSFC under Grants No.12474287, No.12174436 and No.T2121001.
\end{acknowledgments}
\appendix
\section{The specific form of $\mathbf{M}$ matrix}
Here we show the explicit form of the matrix $\mathbf{M}$ with different boundary dissipators. For the dissipative form under consideration, there are four possible combinations for the matrix $\mathbf{M}_{l/g,l/g}$ in Eq.(\ref{MM}), represented as:
\begin{align}
\mathbf{M}_{l/g,l/g}=\left(\begin{array}{ccccc}
\mathbf{G}_{l/g}(\gamma_l) & 0 &  &  & \\
 0 & 0 & \\
 & \ddots & \ddots & \ddots\\
 &  &  & 0 & 0\\
 &  &  & 0 & \mathbf{G}_{l/g}(\gamma_r)
\end{array}\right),\label{M1}
\end{align}
with
\begin{align}
\mathbf{G}_{l/g}(\gamma)=\left(\begin{array}{cc}
\frac{1}{2}\gamma & \mp\frac{\mathrm{i}}{2}\gamma  \\
\pm\frac{\mathrm{i}}{2}\gamma & \frac{1}{2}\gamma
\end{array}\right).
\end{align}
Here the subscript of $\mathbf{M}_{l/g,l/g}$ means that the dissipator is loss or gain on left and right boundary, respectively.

From Eq.(\ref{M1}), we see that, for different combinations, the real part of them are same:
\begin{equation}
\mathbf{M}_{\mathbf{R}}=\Re[\mathbf{M}_{l,l}]=\Re[\mathbf{M}_{l,g}]=\Re[\mathbf{M}_{g,l}]
=\Re[\mathbf{M}_{g,g}].
\end{equation}
Thus, the spectrum matrix $\mathbf{X}=-2 \mathrm{i}\mathbf{H}+2 \mathbf{M}_\mathbf{R}$ is identical for all combinations. Consequently, the Liouvillian spectrum remains unchanged.

The difference among these matrices lies in the imaginary part. While the imaginary part does not influence the Liouvillian spectrum, it plays a crucial role in determining the steady state in Eq.({\ref{eq:lyapunov}) and Eq.(\ref{eq:xy}).

\section{The exact solution of matrix $\mathbf{P}$}
Here we focus on the rapidity spectra, which can be analytically
derived by solving the eigenvalues of the matrix $\mathbf{P}$. We consider the general
case with $t_{1}\neq0$ and $t_{2}\neq0$ and solve the eigenvalue equation
\begin{equation}
\mathbf{P}(t_{1}, t_{2}, \gamma_{l}, \gamma_{r})|\Psi\rangle=E_p|\Psi\rangle\label{Peq}
\end{equation}
by following the analytical method in Ref.\cite{zheng2023prb,GuoCX}, where we
denote $|\Psi\rangle=\left(\psi_{1,A},\psi_{1,B},\psi_{2,A},\cdots,\psi_{N,B}\right)^{T}$. By substituting Eq.(\ref{P}) into Eq.(\ref{Peq}), we get of a series
of bulk equations
\begin{equation}
\begin{array}{c}
t_{2}\psi_{(n-1),B}+t_{1}\psi_{n,B}-E_p\psi_{n,A}=0,\\
t_{1}\psi_{(n-1),A}+t_{2}\psi_{n,A}-E_p\psi_{(n-1),B}=0,
\end{array}\label{eq:Ebulk}
\end{equation}
with $n=2,\cdots,N$, and the boundary equations given by
\begin{equation}
\begin{array}{c}
\mathrm{i}\gamma_{l}\psi_{1,A}+t_{1}\psi_{1,B}-E_p\psi_{1,A}=0,\\
t_{1}\psi_{N,A}+\mathrm{i}\gamma_{r}\psi_{N,B}-E_p\psi_{N,B}=0.
\end{array}\label{eq:Eboundry}
\end{equation}
Here, we set the ansatz of wave function $|\Psi_{P}\rangle$ as follows,
\begin{equation}
|\Psi\rangle=\left(z\phi_{A},z\phi_{B},z^{2}\phi_{A},z^{2}\phi_{B},\cdots,z^{N}\phi_{A},z^{N}\phi_{B}\right)^{T}.\label{waveF}
\end{equation}
where $z=e^{i\theta}$ or $z=e^{-i\theta}$.

By applying the analytical method in Refs.\cite{GuoCX,zheng2023prb,Alase},  the eigenvalue can be represented as
\begin{equation}
E_p=\pm\sqrt{t_{1}^{2}+t_{2}^{2}+2t_{1}t_{2}\cos\theta}.\label{BE}
\end{equation}
The value of $\theta$ is determined by the following equation:
\begin{equation}
p_{1}\sin[N\theta]-p_{2}\sin[(N+1)\theta]+p_{3}\sin[(N-1)\theta]=0,\label{theta}
\end{equation}
where $p_{1}=\mathrm{i}t_{2}(\gamma_{l}+\gamma_{r})E_p-(t_{2}^{3}-t_{2}\gamma_{l}\gamma_{r})$,
$p_{2}=t_{1}t_{2}^{2}$, and $p_{3}=t_{1}\gamma_{l}\gamma_{r}$.

For Eq.(\ref{theta}), the two sets of parameters, $(t_{1}, t_{2}, \gamma_{l}, \gamma_{r})$ and $(-t_{1}, -t_{2}, \gamma_{l}, \gamma_{r})$, yield identical solutions, resulting in $E$ and $E'$ being equivalent in Eq.(\ref{BE}).

\section{The calculation details of single-particle correlation function}
In this appendix, we give the details for the calculation of evolution of particle number density given by
\begin{equation}\label{apeq:spc}
 n(t)=\frac{1}{2N}\sum_{i=1}^{2N}\text{Tr}[c_{i}^{\dagger}c_{i}\rho(t)].
\end{equation}
For convenience, we calculate by the representation of Majorana fermion with $\bar{w}_{2j-1}=\frac{i}{\sqrt{2}}(c_{j}-c_{j}^{\dagger}); \bar{w}_{2j}=\frac{1}{\sqrt{2}}(c_{j}+c_{j}^{\dagger})$ and define the correlation matrix $\Gamma$ in Majorana fermion $\Gamma_{j,k}=i\left\langle \bar{w}_{j}\bar{w}_{k}\right\rangle -\frac{i}{2}\delta_{j,k}$. Then, we can get the Lyapunov equation:
\begin{equation}
\partial_{t}\Gamma(t)=\bar{\mathbf{X}}\Gamma(t)+\Gamma(t) \bar{\mathbf{X}}^{T}+\bar{\mathbf{Y}}.\label{eq:lyapunov}
\end{equation}
The matrix $\bar{\mathbf{X}}$, $\bar{\mathbf{Y}}$ are represented in terms of $2N\times 2N$ non-hermitian matrix which are defined as
\begin{equation}\label{eq:xy}
  \begin{aligned}
  \bar{\mathbf{X}}&=-2i\bar{\mathbf{H}}-2\Re(\sum_{\mu}\bar{l}_{\mu}\bar{l}_{\mu}^{\dagger}) ,\\ \bar{\mathbf{Y}}&=2\Im(\sum_{\mu}\bar{l}_{\mu}\bar{l}_{\mu}^{\dagger}),
  \end{aligned}
\end{equation}
where $H =\sum_{j, k=1}^{N} \bar{w}_j \bar{\mathbf{H}}_{j, k} \bar{w}_k; L_\mu =\sum_{j=1}^{N} \bar{l}_{\mu, j} \bar{w}_j$.

Here, we mainly consider about the dynamics, especially the behaviors of converging to the steady state. So, we can calculate only the deviation $\bar{\Gamma}(t)=\Gamma(t)-\Gamma_{s}$, where $\Gamma_{s}(t)$ means the matrix form of the steady state. And the evolution of $\hat{\Gamma}(t)$ is determined by
\begin{equation}\label{eq:hatGamma}
  \bar{\Gamma}(t)=e^{\bar{\mathbf{X}} t} \bar{\Gamma}(0) e^{\mathbf{X}^{T} t}.
\end{equation}
By using the relation $\Gamma(t)=\bar{\Gamma}(t)+\Gamma_{ss}$, we can get
\begin{equation}
\label{cor}
\begin{aligned}
n(t)&=\frac{1}{2N}\sum_{j=1}^{2N}i\langle \bar{w}_{2j-1}\bar{w}_{2j}\rangle+\frac{1}{2}\\
&=\frac{1}{2}+\frac{1}{2N}\sum_{j=1}^{2N}\Gamma_{2j-1,2j}.
\end{aligned}
\end{equation}

\section{Analytic analysis of dynamical duality from the rapidity spectrum}
The rapidity spectrum is obtained by solving Eq.~(\ref{theta}). For clarity and without loss of generality, we primarily focus on the symmetric case with $\gamma_{l} = \gamma_{r} = \gamma$ to analyze the duality in the rapidity spectrum and we shall set $t_{2}=1$ as the energy unit. The explicit form of Eq.~(\ref{theta}) is given below:
\begin{equation}\label{appdx-e}
\begin{aligned}
(\gamma^{2}- & 1+2\mathrm{i}\gamma E_{p})\sin[N\theta]+t_{1}\gamma^{2}\sin[(N-1)\theta]
\\&-t_{1}\sin[(N+1)\theta]=0,
\end{aligned}
\end{equation}
Corresponding to $E_{\pm}(\theta_{\pm})=\pm \sqrt{1+t_{1}^{2}+2t_{1}\cos[\theta_{\pm}]}$, here we denote the corresponding solutions as $\theta_{\pm}$, respectively, i.e.,
\begin{equation}\label{appdx-e1}
\begin{aligned}
(\gamma^{2}- & 1+2\mathrm{i}\gamma E_{\pm}(\theta_{\pm}))\sin[N\theta_{\pm}]+t_{1}\gamma^{2}\sin[(N-1)\theta_{\pm}]
\\&-t_{1}\sin[(N+1)\theta_{\pm}]=0.
\end{aligned}
\end{equation}
According to Eq.~(\ref{appdx-e1}), assuming $\theta_{+}$ is a solution of the equation associated with the eigenvalue $E_{+}(\theta_{+})$, then $\theta_{+}^{*}$ is also a solution, corresponding to the eigenvalue $E_{-}(\theta_{-})$. This implies the relation $\theta_{-}(\lambda) = \theta_{+}^{*}(\lambda)$. Similarly, one can show that $\theta_{+}(\lambda) = \theta_{-}^{*}(\lambda)$. As a consequence, the eigenvalues satisfy the relations
$
E_{-}(\theta_{-}) = E_{-}(\theta_{+}^{*}) = -E_{+}(\theta_{+}^{*}) = -\left( E_{+}(\theta_{+}) \right)^*
\quad \text{and} \quad
E_{+}(\theta_{+}) = -E_{-}(\theta_{-}^{*}) = -\left( E_{-}(\theta_{-}) \right)^*.
$
In the following, we first focus on the case of $E_{+}$ and derive the corresponding equation as:
\begin{equation}\label{appdx-e2}
\begin{aligned}
(\gamma^{2}- & 1+2\mathrm{i}\gamma  E_{+}(\theta_{+}))\sin[N\theta_{+}]+t_{1}\gamma^{2}\sin[(N-1)\theta_{+}]
\\&-t_{1}\sin[(N+1)\theta_{+}]=0.
\end{aligned}
\end{equation}
Next, we focus on the eigenvalue equation for $ E_{-} $ with dissipation strength $ \gamma' $, and Eq.~(\ref{appdx-e1}) can be rewritten as
\begin{equation}\label{appdx-e3}
\begin{aligned}
(\gamma'^{2}- & 1 + 2\mathrm{i} \gamma' E_{-}(\theta_{-})) \sin(N\theta_{-})
+ t_{1} \gamma'^{2} \sin[(N-1)\theta_{-}]
\\&- t_{1} \sin[(N+1)\theta_{-}] = 0.
\end{aligned}
\end{equation}
Using the relations $ E_{-}(\theta_{-}) = -\left( E_{+}(\theta_{+}) \right)^* = -E_{+}(\theta_{+}^*) $ and $ \theta_{-} = \theta_{+}^* $, Eq.~(\ref{appdx-e3}) becomes
\begin{equation}
\begin{aligned}
(\gamma'^{2} - & 1 - 2\mathrm{i} \gamma' E_{+}(\theta_{+}^*)) \sin(N\theta_{+}^*)
+ t_{1} \gamma'^{2} \sin[(N-1)\theta_{+}^*]
\\&- t_{1} \sin[(N+1)\theta_{+}^*] = 0.
\end{aligned}
\end{equation}
Now, assuming $\gamma' = 1/\gamma $, the equation becomes
\begin{equation}\label{appdx-e6}
\begin{aligned}
(\gamma^{2} - & 1 + 2\mathrm{i} \gamma E_{+}(\theta_{+}^*)) \sin(N\theta_{+}^*)
+ t_{1} \gamma^{2} \sin[(N+1)\theta_{+}^*]
\\&- t_{1} \sin[(N-1)\theta_{+}^*] = 0.
\end{aligned}
\end{equation}
For bulk states, the real part of $ \theta_{+} $ scales as $ 1/N $ and the imaginary part scales as $ 1/N^2 $. Therefore, when $N$ is large enough, we can approximate
$
\sin[(N-1)\theta_{+}] \approx \sin[(N+1)\theta_{+}].
$
Under this approximation, Eq.~(\ref{appdx-e6}) becomes equivalent to Eq.~(\ref{appdx-e2}). Since Eqs.~(\ref{appdx-e3}) and (\ref{appdx-e6}) are equivalent under the transformation $ \gamma' = 1/\gamma $, we conclude that
$
\theta_{+}(\gamma) \approx \theta_{-}^*\left(\tfrac{1}{\gamma}\right)
$
in the thermodynamic limit. Using the additional relation $ \theta_{-}^*\left(\tfrac{1}{\gamma}\right) \approx \theta_{+}\left(\tfrac{1}{\gamma}\right) $, we obtain
$
\theta_{+}(\gamma) \approx \theta_{+}\left(\tfrac{1}{\gamma}\right),
$
and consequently,
$
E_{+}(\gamma) \approx E_{+}\left(\tfrac{1}{\gamma}\right).
$
A similar argument applies for \( E_{-} \), yielding
$
E_{p}(\gamma) \approx E_{p}\left(\tfrac{1}{\gamma}\right)
$
in the thermodynamic limit.
We emphasize that this approximation does not hold for boundary-bound states, as the real part of $ \theta_{+} $ in those cases is proportional to $ N^{0} $, and thus the approximation $ \sin[(N-1)\theta_{+}] \approx \sin[(N+1)\theta_{+}] $ breaks down.

\section{The exact solution of the dark state}
In this appendix, we analyze analytically the condition for the occurrence of the dark state in the thermodynamic limit. Since we consider the case with only a single boundary dissipator, we take $\gamma_{l}=0, \gamma_{r}=\gamma$ or $\gamma_{r}=0, \gamma_{l}=\gamma$ and assume the solution of Eq.(\ref{theta}) as $\theta=\theta_{R}+\mathrm{i}\theta_{I}$ and $\theta_{R}$, $\theta_{I}$ are purely real ($\theta_{R} \in [0,2\pi)$). Then, we can rewrite the corresponding eigenvalue as
\begin{equation}
E_p =\pm\sqrt{1+t_{1}^{2}+2t_{1}\cos[\theta_{R}+\mathrm{i}\theta_{I}]}.\label{AE1}
\end{equation}
The eigenvalue $E$ equal to $0$ which corresponds to the dark state, thus we should have $\theta_{R}=\pi$, and the eigenvalue can be expressed as
\begin{equation}
E_p=\pm\sqrt{1+t_{1}^{2}-2t_{1}\cosh[\theta_{I}]}.\label{AE2}
\end{equation}

Firstly, in the case where $\theta_{I}=0$, we can determine that when $t_{1}=1$, the energy $E_p=0$ is satisfied. Then, we consider the case $\theta_{I}\neq0$. Substituting $\theta = \pi + \mathrm{i}\theta_{I}$ into Eq.(\ref{theta}) , we can get
\begin{equation}\label{Athetaeq}
\begin{aligned}
\gamma^{2} E_p^{2} \sinh[N \theta_{I}]^2+[\sinh[N \theta_{I}]-t_{1}\sinh[(N+1)\theta_{I}]]^{2}=0.
\end{aligned}
\end{equation}
In the thermodynamic limit of $N \rightarrow \infty$, assuming $\theta_{I}>0$, we can obtain the approximate equation:
\begin{equation}
2\sinh\left[N\theta_{I}\right]\thickapprox e^{N \theta_{I}},
\end{equation}
and thus the Eq.(\ref{Athetaeq}) is simplified to
\begin{equation}
(1-t_{1}e^{\theta_{I}})^2+\gamma^2(t_{1}^2+1-2 t_{1} \cosh[\theta_{I}])=0,
\end{equation}
which gives rise to
\begin{equation}
e^{-\theta_{I}}(t_{1}e^{\theta_{I}}-1)(t_{1}e^{2\theta_{I}}-e^{\theta_{I}}(1+\gamma^{2})
t_{1}\gamma^2)=0 .
\end{equation}
Let $x=e^{\theta_{I}}$, and the solutions of $x$ are given by
\begin{align}
x_{\pm}= \frac{1+\gamma^{2}\pm\sqrt{1+\gamma^{4}+\gamma^{2}(2-4t_{1}^{2})}}{2t_{1}}, x=\frac{1}{t_{1}}.
\end{align}

Substituting the solutions into Eq.(\ref{AE2}), we find that only the solution $x=\frac{1}{t_{1}}$ yields spectra satisfying $E_p=0$, corresponding to the dark state. Cause $x=e^{\theta_{I}}>1$, thus we can get that the condition for the appearance of the dark state is $0<t_{1}<1$ for $\theta_{I}>0$. Based on the discussion above, we can get the appearance of the dark state is $0<t_{1}\leq1$ corresponding to the topologically nontrivial region. In the topologically nontrivial region, the solutions $x_{\pm}$ satisfy $x_{+}>1, x_{-}<1$. Therefore, the solution $x_{-}$ is discarded, while the solution $x_{+}$ corresponds to the bound state.
If assuming $\theta_{I}<0$, we would arrive at the same conclusion.

Since $x=\frac{1}{t_{1}}$ is independent of the dissipation strength, the distribution of the dark state wave function is also independent of the dissipation strength but depends on hopping amplitudes between intra-cell and inter-cell sites.

\section{Dynamics in the presence of boundary defects}
In the topological phase, localized modes exist, and we observe dynamical duality behavior. It is interesting to ask whether dynamical duality behavior can also emerge in non-topological systems when localized states are present? To illustrate this, we consider
a simple model where the Hamiltonian is taken as
\begin{equation}
H=\sum_{j=1}^{N}J(c_{j}^{\dagger}c_{j+1}+c_{j+1}^{\dagger}c_{j})
+V(c_{1}^{\dagger}c_{1}+c_{L}^{\text{\ensuremath{\dagger}}}c_{L}).
\end{equation}
For convenience, we set $J=1$. It can be readily verified that, in this model, localized edge states emerge when the strength of boundary defects exceeds a threshold $V>1$, whereas the localized edge states are absent when $V<1$.
The dissipation operators are chosen as $L_{l}=\sqrt{\gamma_{l}}c_{1}$ and $L_{r}=\sqrt{\gamma_{r}}c_{L}$ in following discussion.

\begin{figure}[h]
\centering \includegraphics[width=8.5cm]{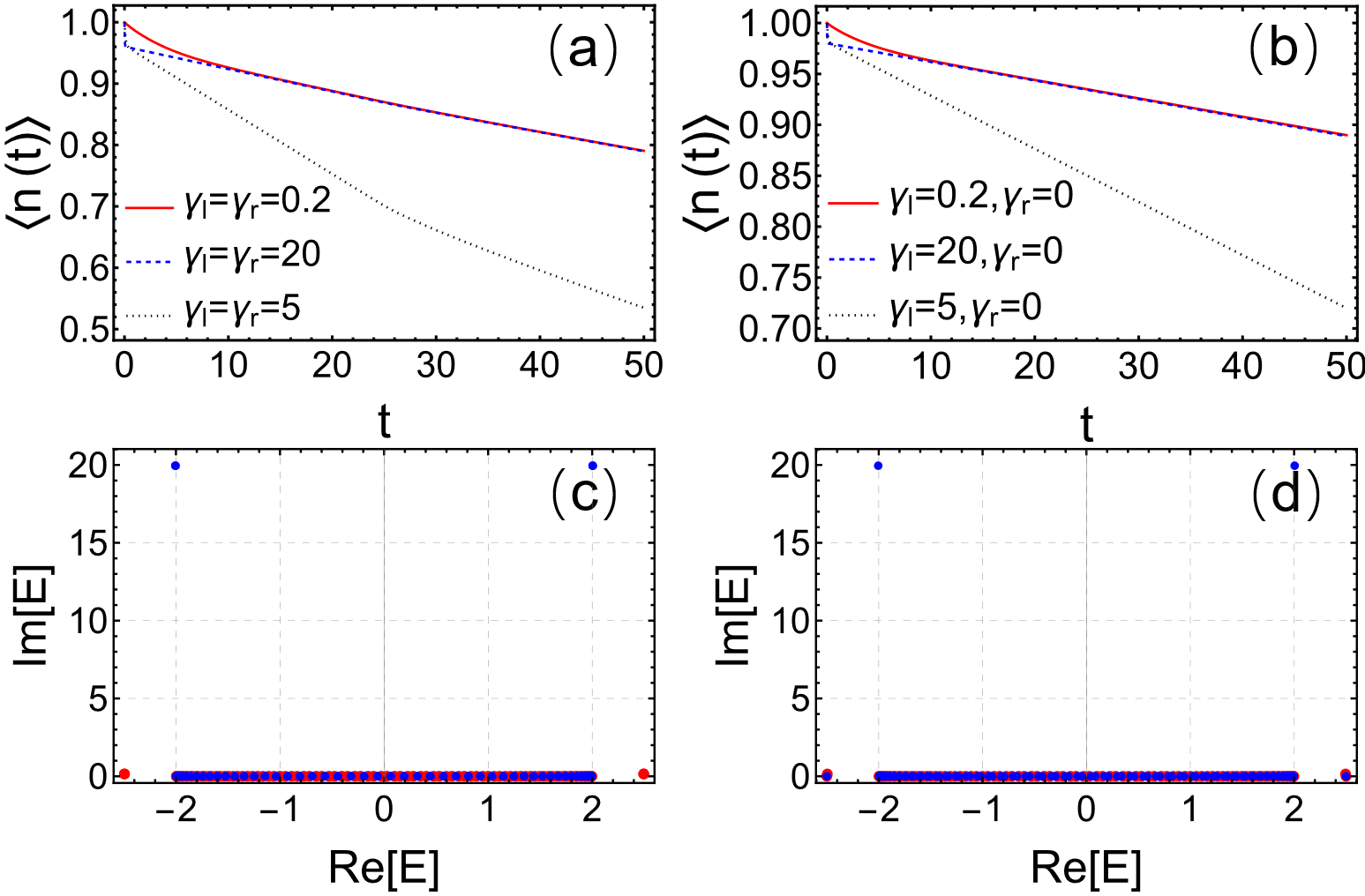}
\caption{Panels (a) and (b): Time evolution of the particle number density for $N = 50$, $J = 1$, and $V = 2$. Panels (c) and (d): Rapidity spectra with $N = 50$, $J = 1$, and $V = 2$. In panel (c), the red points correspond to $\gamma_{l} = \gamma_{r} = 0.2$, while the blue points correspond to $\gamma_{l} = \gamma_{r} = 20$. In panel (d), the red points represent the case of $\gamma_{l} = 0.2$, $\gamma_{r} = 0$, and the blue points correspond to $\gamma_{l} = 20$, $\gamma_{r} = 0$.}
\label{fig8}
\end{figure}

We investigate two representative cases: one with $\gamma_{l} = \gamma_{r}=\gamma$ and the other with $\gamma_{l} =\gamma$ and $\gamma_{r} = 0$. In both cases, when localized states are present ($V>1$), we find the existence of dynamical duality in the long time dynamical evolution when the dissipation strengths fulfills the following relation
\begin{equation}
\gamma' = V^2 / \gamma.
\end{equation}
Next we demonstrate the dynamical evolution of the particle number density  in Figs.\ref{fig8}(a) and (b). We observe that the long-time dynamics indeed displays a clear signature of dynamical duality for $\gamma = 0.2$ and $\gamma = 20$.  As a contrast, we also display the data for $\gamma = 0.2$ and $\gamma = 5$ (fulfilling $\gamma' = 1 / \gamma$). In this case, the dynamical behaviors in the strong and weak dissipation regions deviate significantly. For the case of $V<1$, we do not observe behaviors of dynamical duality. 

To further understand this dual behavior, we also check the corresponding rapidity spectra for systems with $\gamma = 0.2$ and $\gamma = 20$. Specifically, all modes except for the boundary-bound states show similar behavior, and the number of boundary-bound states is also identical. As shown in Figs.\ref{fig8}(c) and (d), we choose two sets of dissipation strengths satisfying the dynamical duality condition, namely $\gamma = 0.2$ and $\gamma = 20$. The resulting rapidity spectra exhibit similar structures, except at the points associated with localized modes. This indicates that, the Liouvillian modes closest to the steady state remain essentially unchanged between the weak and strong dissipation limits and this leads to dynamical duality.


\end{document}